\documentclass[prl,aps,twocolumn,groupedaddress,showpacs]{revtex4}

\usepackage{bm}
\usepackage{graphicx}

\begin{document}

\title{Reconstruction of Band Structure Induced by Electronic Nematicity in an FeSe Superconductor}

\author{K. Nakayama,$^1$ Y. Miyata,$^1$ G. N. Phan,$^1$ T. Sato,$^1$ Y. Tanabe,$^1$ T. Urata,$^1$ K. Tanigaki,$^{1,2}$ and T. Takahashi$^{1,2}$}

\affiliation{$^1$Department of Physics, Tohoku University, Sendai 980-8578, Japan\\
$^2$WPI Research Center, Advanced Institute for Materials Research, Tohoku University, Sendai 980-8577, Japan
}

\date{\today}

\begin{abstract}
We have performed high-resolution angle-resolved photoemission spectroscopy on FeSe superconductor ($T_c$ $\sim$ 8 K), which exhibits a tetragonal-to-orthorhombic structural transition at $T_s$ $\sim$ 90 K. At low temperature we found splitting of the energy bands as large as 50 meV at the M point in the Brillouin zone, likely caused by the formation of electronically driven nematic states. This band splitting persists up to {\it T} $\sim$ 110 K, slightly above $T_s$, suggesting that the structural transition is triggered by the electronic nematicity. We have also revealed that at low temperature the band splitting gives rise to a van Hove singularity within 5 meV of the Fermi energy. The present result strongly suggests that this unusual electronic state is responsible for the unconventional superconductivity in FeSe.
\end{abstract}


\maketitle
Iron-based superconductors (Fe SCs) have a rich phase diagram \cite{Stewart}, wherein most parent compounds exhibit a tetragonal-to-orthorhombic structural transition, as well as a collinear-type antiferromagnetic transition. These two transitions are typically strongly coupled, leading to identical or very similar transition temperatures. Superconductivity generally arises when these transitions are suppressed by doping carriers or applying pressure, leading to a characteristic superconducting dome in the electronic phase diagram. Recently, evidence has been mounting for the existence of other exotic states in the phase diagram called {\it nematic} states \cite{FernandesReview, Chu, Tanatar, Chuang, Shen122, Zhao, Kasahara, Shimojima, Feng, Shen111, Rosenthal}. These states have been reported in a variety of systems including quantum Hall states, ruthenium oxides, and high-$T_{\rm c }$ copper oxides \cite{Fradkin}. In the nematic states of Fe SCs, the tetragonal (C$_4$) rotational symmetry of the Fe plane is spontaneously broken. Intensive experimental investigation in the orthorhombic phase of the {\it 122} system {\it A}Fe$_2$As$_2$ ({\it A} = Ba, Sr, Ca) has revealed a strong in-plane anisotropy possessing C$_2$ symmetry in transport measurements, electronic states, and magnetic excitations, indicative of nematicity \cite{Chu, Tanatar, Chuang, Shen122, Zhao}. Nematicity has also been reported in the tetragonal phase of BaFe$_2$(As,P)$_2$ \cite{Kasahara, Shimojima} and in the {\it 111} NaFeAs \cite{Feng, Shen111, Rosenthal} system. While these studies have provided important insight into the unconventional nematic states, it is still unclear whether the nematicity observed in these two categories of Fe SCs is a fundamental phenomenon among all Fe SCs, and whether the nematicity is related to the emergence of superconductivity.

Bulk FeSe (the {\it 11} system) offers an excellent opportunity to resolve above issue, since it exhibits a tetragonal-to-orthorhombic transition at $T_s$ $\sim$ 90 K {\it without} long-range magnetic order \cite{Hsu, McQueen} that might complicate the electronic states \cite{Shen122, Feng, Shen111, PierreDirac}. While most Fe SCs show superconductivity in the tetragonal phase, superconductivity in FeSe ($T_c$ $\sim$ 8 K) emerges in the orthorhombic phase. FeSe has also attracted considerable attention because of the discovery of superconductivity around the boiling point of liquid nitrogen in monolayer FeSe on SrTiO$_3$ \cite{XueSTO}. While FeSe is certainly unique among the Fe SCs, few experimental studies on the electronic states of FeSe have been performed \cite{Xue, BorisenkoFeSe}, largely due to the difficulty of growing high-quality single crystals. However, recent breakthroughs to grow bulk FeSe single crystals \cite{Bohmer} enable the fabrication of crystals suitable for angle-resolved photoemission spectroscopy (ARPES) measurements.

In this Letter, we report ARPES results of high-quality FeSe single crystals ($T_c$ $\sim$ 8 K)  \cite{Tanabe}. We have revealed that the electronic structure undergoes a considerable reconstruction as a function of temperature. We also found evidence for nematic electronic states which develop slightly above $T_s$, as seen in the lifting of the band degeneracy around the M point and the change in the band dispersion around the $\Gamma$ point. Based on these results, we discuss the interplay between the nematicity, magnetic order, and superconductivity.

High-quality single-crystals of FeSe were grown by the KCl and AlCl$_3$ flux method \cite{Bohmer,Lin,Chareev}. Details of the sample preparation are described elsewhere \cite{Tanabe}. High-resolution ARPES measurements were performed at Tohoku University using a VG-Scienta SES2002 spectrometer and a He discharge lamp ($h\nu$ = 21.218 eV). Additional ARPES measurements were performed with synchrotron radiation at BL-28A at Photon Factory (KEK) with a VG-Scienta SES2002 spectrometer using circularly polarized 40-eV photons. The energy and angular resolutions were set at 12-30 meV and 0.2$^{\circ}$, respectively. Clean sample surfaces were obtained by cleaving crystals {\it in-situ} in ultrahigh vacuums better than 1$\times$10$^{-10}$ Torr. The Fermi level $E_{\rm F}$ of the samples was referenced to that of a gold film evaporated onto the sample holder.
\begin{figure}
\includegraphics[width=3.4in]{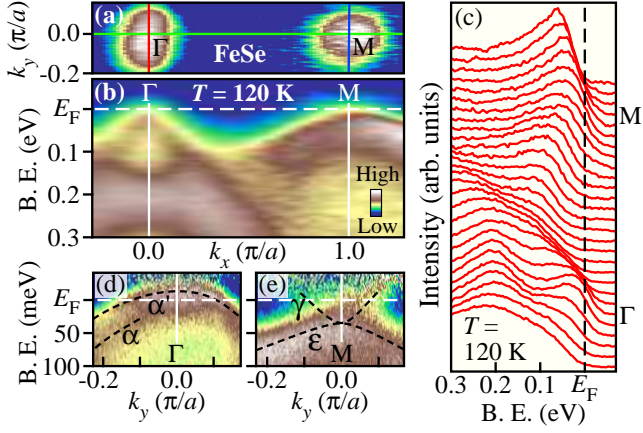}
\vspace{-0.5cm}
\caption{(a) ARPES-intensity mapping for FeSe at $E_{\rm F}$ in a 2D wave-vector plane around the $\Gamma$-M cut obtained with He-I$\alpha$ photons ($h\nu = 21.218$ eV). The map is obtained by integrating the spectral intensity within $\pm$5 meV of $E_{\rm F}$. (b) (c) The ARPES intensity and corresponding EDCs, respectively, measured along the green line in (a). (d) (e) Near-$E_{\rm F}$ ARPES intensity along the red and blue lines, respectively, in (a) divided by the Fermi-Dirac distribution function for $T$ = 120 K convoluted with the instrumental resolution. Dashed curves are a guide for eyes.}
\end{figure}

First, we present the electronic states above $T_s$ of FeSe. Figure 1(a) shows the Fermi surface (FS) for FeSe around the $\Gamma$-M cut of the Brilliron zone (BZ) at $T$ = 120 K. Two high intensity spots centered at the $\Gamma$ and M points are clearly visible, corresponding to the two kinds of FSs typically observed in Fe SCs such as FeTe$_{1-x}$Se$_x$ ($x \leq 0.5$) \cite{Nakayama, PierreReview}. Figures 1(b) and 1(c) show the ARPES intensity plot and energy distribution curves (EDCs) along the  $\Gamma$-M line. We observe a highly dispersive holelike band at the $\Gamma$ point and a less-dispersive holelike band around the M point. The band at the $\Gamma$ point consists of two branches as seen in Fig. 1(d) [also see the second-derivative intensity plot in Fig. 2(e)]. One branch, referred as the $\alpha'$ band, crosses $E_{\rm F}$ and reaches a maximum energy of $\sim$10 meV above $E_{\rm F}$. The other branch, $\alpha$ band, has a binding energy 20-40 meV higher than that of $\alpha$ band. According to a previous ARPES study \cite{BorisenkoFeSe}, these bands originate from the Fe 3$d_{zx}/d_{yz}$ orbitals. Examining the ARPES intensity around the M point [Fig. 1(e)], the holelike ($\epsilon$) band has a maximum at $\sim$40 meV below $E_{\rm F}$, and connects to another weaker electronlike band ($\gamma$ band) at the M point, consistent with the band calculations \cite{Subedi}. These observations establish that the basic FS topology in the tetragonal phase is universal in FeTe$_{1-x}$Se$_x$ regardless of the Se content.

At low temperature, we observe a drastic reconstruction of the band structure. Figures 2(a) and 2(b) show the ARPES intensity and corresponding EDCs, respectively, along the $\Gamma$-M cut at $T$ = 30 K. Comparing the result at 30 K with that at 120 K, the existence of two holelike bands at the M point at $T$ = 30 K [black dashed curves in Fig. 2(a)] contrasts with the single holelike band seen at $T$ = 120 K [Fig. 1(b)]. This difference is also visible from the single- {\it vs}. two-peaked shape of the EDCs near the M point [Figs. 1(c) and 2(b)]. By referring to previous ARPES studies on BaFe$_2$As$_2$ and NaFeAs \cite{Shen122, Feng, Shen111}, we suggest that the two-peaked structure at the M point originates from an anisotropic energy shift of the $d_{zx}$ and $d_{yz}$ orbitals, reflecting the development of nematic electronic states below $T_s$ \cite{Shen122, Feng, Shen111}. It has been reported that the holelike band with the dominant $d_{yz}$ character shifts upward along the (0, 0)-($\pi$, 0) direction of the untwinned crystal, while the holelike band with  dominant $d_{zx}$ character shifts downward along the (0, 0)-(0, $\pi$) direction [see Fig. 2(c) and Refs. [6, 10, 11], leading to the emergence of C$_2$-symmetric electronic states. In our experiment, these two bands are simultaneously observed around the M point since in the twinned FeSe crystal, the (0, 0)-($\pi$, 0) and (0, 0)-(0, $\pi$) directions of the untwinned crystal are inherently mixed in both $k_x$ and $k_y$ directions. In this regard, the observation of a single peak in the EDCs at $T$ = 120 K [Fig. 1(c)] is quite natural since the $d_{yz}$ and $d_{zx}$ orbitals become degenerate at the M point due to the C$_4$ symmetry of the crystal.

\begin{figure}
\includegraphics[width=3.4in]{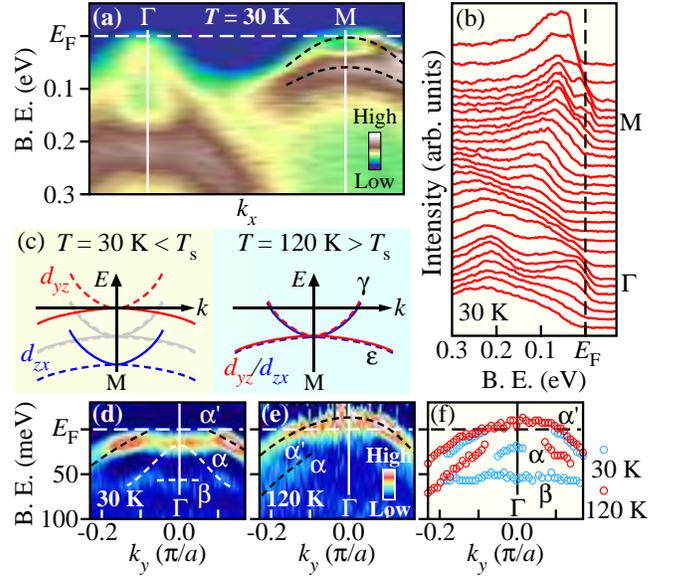}
 \vspace{-0.5cm}
 \caption{(a),(b) ARPES intensity and corresponding EDCs, respectively, measured along the $\Gamma$-M direction. Dashed curves in (a)  trace the M-centered holelike bands. (c) Schematic band diagram around the M point below and above $T_s$. Red and blue curves indicate the $d_{yz}$ and $d_{zx}$ orbitals. Solid and dashed curves depict the band dispersion along the (0, 0)-($\pi$, 0) and (0, 0)-(0, $\pi$) directions (long and short Fe-Fe directions) of the untwinned crystal, respectively. (d) (e) The second-derivative plot of the near-$E_{\rm F}$ ARPES intensity around the $\Gamma$ point for $T$ = 30 and 120 K. (f) Temperature dependence of band dispersion around the $\Gamma$ point, extracted by tracing the peak maxima of the EDCs divided by the Fermi-Dirac distribution function.}
\end{figure}

\begin{figure*}
\includegraphics[width=6.4in]{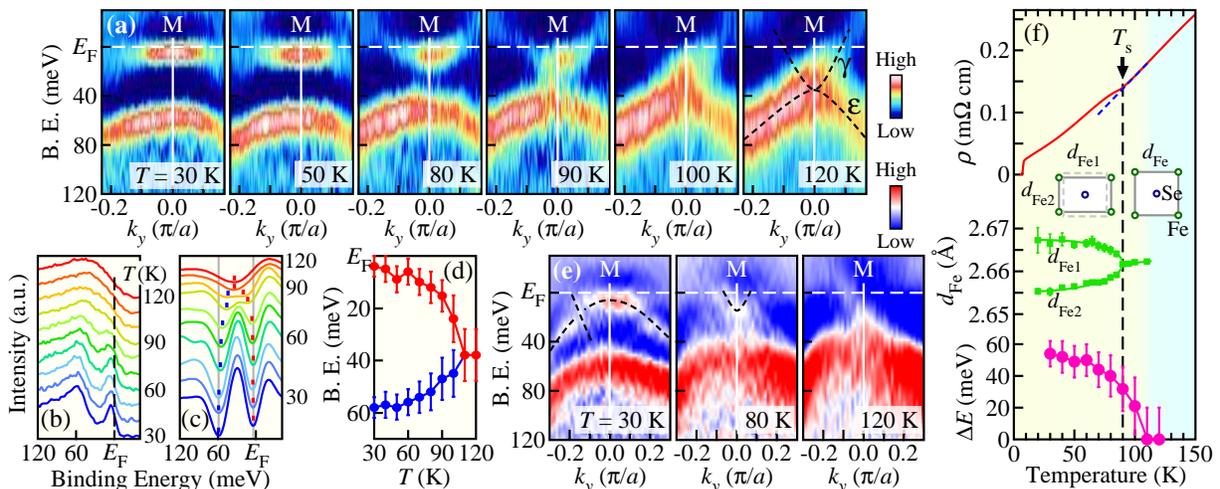}
 \vspace{-0.5cm}
 \caption{(a) Second-derivative plot of the near-$E_{\rm F}$ ARPES intensity around the M point taken along the {\bf k} cut shown by blue line in Fig. 1(a) measured using the the He I$\alpha$ line, at various temperatures. (b) (c), Temperature dependence of the EDC at the M point and its second derivative, respectively. Blue and red dots in (c) indicate the local minima corresponding to the peak position in (b). (d) Temperature dependence of the peak energies in the EDC at the M point [same as dots in (c)]. (e) Same as (a) but measured with circularly polarized 40-eV photons. In addition to the high-energy holelike band seen in (a), other M-centered holelike and electronlike bands are observed at {\it T} = 30 K, as indicated by the dashed curves. The dashed curve at {\it T} = 80 K is a guide for eyes, tracing the near-$E_{\rm F}$ electronlike band (this band is indistinguishable at {\it T} = 30 K due to the band degeneracy with the low-energy holelike band). (f) Comparison of the temperature dependence of the electrical resistivity recorded from the same sample used for the ARPES (red curve) \cite{Tanabe}, the Fe-Fe distance estimated from the X-ray diffraction (green squares) \cite{McQueen}, and the magnitude of the band splitting at the M point (purple circles). The blue dashed line in the resistivity plot highlights the anomaly above $T_s$.}
\end{figure*}

In addition to the band reconstruction around the M point, a characteristic change is also observed at the $\Gamma$ point. As shown in the second-derivative plot of the ARPES intensity at $T$ = 30 K in Fig. 2(d), a bright spot is observed at $\sim$20 meV below $E_{\rm F}$. This is ascribed to the top of the $\alpha$ band due to its holelike character. If we examine the dispersion of this band empirically, we also see a strong similarity to the spectra at 120 K, as plotted in Fig. 2(f). Taking into account that the $\alpha'$ band appears energetically stationary across $T_s$, the $\alpha$ and $\alpha'$ bands must then be separated from each other at the $\Gamma$ point at $T$ = 30 K. Such a lifting of the band degeneracy can be explained in terms of the electronic nematicity, where the energy levels of the $d_{zx}$ and $d_{yz}$ orbitals become inequivalent in the orthorhombic phase. Examining Figs. 2(d)-2(f), one also observes a relatively flat band at $\sim$60 meV below $E_{\rm F}$ ($\beta$ band) only for $T$ = 30 K, which should then have an orbital character different from the $\alpha$ and $\alpha'$ bands.

To clarify the relationship between the changes in the band dispersion and the structural transition, a systematic temperature-dependent ARPES measurement was also performed. Since the band splitting at the M point is related to the strength of the nematicity, we chose a {\bf k} cut which crosses the M point. As shown in Fig. 3(a),  at $T$ = 30 K we find distinct high intensity distributions near $E_{\rm F}$ and $\sim$60 meV arising from the energy difference between the $d_{zx}$ and $d_{yz}$ orbitals. At $T$ = 90 K, thanks to a downward shift of the band as well as the finite population of electrons above $E_{\rm F}$, the near-$E_{\rm F}$ intensity clearly exhibits an electronlike dispersion. This result indicates that the near-$E_{\rm F}$ band observed at low temperatures mainly originates from the bottom of this electronlike band. As the near-$E_{\rm F}$ band dispersion along the $\Gamma$-M cut [green line in Fig. 1(a), taken perpendicular to the cut in Fig. 3(a)] exhibits a holelike character as seen in Fig. 2(a), it is likely that this band has a van Hove singularity at the M point. The raw EDCs show that this singularity point is located within 5 meV of $E_{\rm F}$. As shown in Fig. 3(a), the two-peaked intensity pattern at low temperature gradually smears out above 90 K, and finally becomes invisible at 120 K. The EDC at the M point in Fig. 3(b) further reveals that the two peaks gradually broaden with increasing temperature, and eventually merge into a single peak around $T_s$. We have accurately determined the energy position of the peaks from the local minima of the second derivative of the EDCs [see Fig. 3(c)]. As shown in Fig. 3(d), the energy separation of the two peaks gradually decreases with increasing temperature. Intriguingly, the two peaks appear to merge into a single peak not at $T_s$, but slightly above it ($T$ $\sim$ 110 K).

We attributed the band splitting at the M point to the nematicity. It is worthwhile to consider if this band splitting can be alternatively explained by a ``peak-dip-hump" structure induced by strong electron-phonon coupling as in FeTe \cite{ShenFeTe}. While the ARPES line shape of FeSe resembles that of FeTe, we note two important differences. First, the energy position of the diplike feature in FeSe [see Fig. 3(b); $\sim$30 meV] is $\sim$1.5 times different from that in FeTe (18 meV). This conflicts with the natural expectation that the phonon energies, which reflect the dip energy in the EDCs, should be basically the same in both FeSe and FeTe. Furthermore, the finite energy dispersion around the $\Gamma$ point in FeSe is different from the relatively flat band in FeTe. If strong electron-phonon coupling is essential, a similar flat band should exist around the $\Gamma$ point in FeSe. Thus, electron-phonon coupling cannot account for the band splitting in FeSe. Based on the ARPES spectra it may also seem unusual that only two bands are visible in Fig. 3(a), since in twinned FeSe at low temperatures nemacity should lead to at least four bands at the M point [see Fig. 2(c)]. Their absence is likely due to photoelectron matrix-element effects resulting in suppression of the intensity for two of the four bands. As shown in Fig. 3(e), consistent with our expectations, ARPES measurements with circularly polarized 40-eV photons do reveal additional two bands at $T$ = 30 K; a holelike band whose top is very close to $E_{\rm F}$, and an electronlike band that crosses $E_{\rm F}$ (dashed curves). This confirms the presence of four bands around the M point at low temperature, which merge into two bands at high temperature [see Fig. 3(e)] as shown in Fig. 2(c).

To illustrate the relationship between the nematicity and the structural transition, we plot in Fig. 3(f) the temperature dependence of the electrical resistivity, the Fe-Fe distance ($d_{\rm Fe1}$ or $d_{\rm Fe2}$) \cite{McQueen}, and the magnitude of the band splitting at the M point. It is apparent that $T_s$ and anomaly in resistivity align well with each other. On the other hand, the onset temperature of the band splitting is $\sim$110 K, which is $\sim$20 K higher than $T_s$. This suggests that the nematicity of the electronic states, as inferred from the band splitting, is not a consequence of the structural transition. This conclusion is supported by the observation of a sizable splitting of $\sim$50 meV at low temperatures, which is much larger than the energy-level splitting of $\sim$10 meV determined from band calculations for orthorhombic BaFe$_2$As$_2$, where the distortion should be much larger than that in FeSe (see Supplemental Material of Ref. [6]). It is thus suggested that the observed nematicity is electronic in origin, and is likely a driving force of the structural transition \cite{note}.

Having established that the nematicity is electronically driven, it is important to examine whether the nematicity originates from spin \cite{FernandesReview, spin} or orbital(charge) \cite{orbital} fluctuations. In general these fluctuations are entangled with each other \cite{FernandesReview}, making it difficult to determine the dominant mechanism (spin or orbital) responsible for nematicity. A key finding in this study is the similar degree of band splitting in FeSe and BaFe$_2$As$_2$ \cite{Shen122} in spite of the absence of long-range magnetic order in FeSe. This result supports the orbital-fluctuation scenario. However, the similar onset temperature of the nematicity ($\sim$110 K) and the development of spin fluctuations seen in the NMR measurements \cite{Imai} leaves room for the spin-fluctuation scenario. A theoretical analysis on the splitting size would help to resolve this issue.

Present results also have important implications for understanding superconductivity in FeSe. Because of the electronic nematicity, the underlying electronic structure responsible for superconductivity in FeSe develops the C$_2$ symmetry. This situation rarely occurs in Fe SCs {\it without} the emergence of magnetic order. In the {\it 122} system for example, although the C$_2$ symmetry coexists with superconductivity and magnetic order in the underdoped region \cite{Coexistence}, only a few results for the existence of nematicity have been reported in the optimally doped or overdoped region \cite{Kasahara} (some other results suggested the emergence of nematic fluctuations \cite{Fisher}). The observation of clear band splitting via ARPES has similarly been limited to the underdoped region in the {\it 122} system \cite{Shen122, Shimojima, ShenCoexistence}. Electronic nematicity in FeSe should have a significant impact on the pairing symmetry, as unconventional pairing states have been predicted in the nematic phase \cite{Gap}. Twofold pairing symmetry with nodal lines suggested by previous tunneling-spectroscopy measurements for FeSe \cite{Xue} is likely related to the observed electronic nematicity. The existence of the van Hove singularity around the M point may also affect pairing, as the van Hove singularity near $E_{\rm F}$ is well associated with superconductivity \cite{BorisenkoLiFeAs}. Intriguingly, this condition is well satisfied in FeSe only as a consequence of the lifting of its band degeneracy due to nematicity. It is noted that while several previous studies in the coexistence region of the {\it 122} system revealed competition between nematicity and superconductivity \cite{Coexistence, ShenCoexistence}, a recent x-ray diffraction study has suggested that this is not the case in FeSe \cite{Bohmer}. Therefore it is inferred that the interplay between nematicity and superconductivity in FeSe is different from that in the {\it 122} system. It is important that future work clarifies the relationship between the pairing and the characteristic electronic states of FeSe by accurately determining the {\bf k} dependence of the superconducting gap.

In conclusion, we revealed the development of electronic nematicity slightly above $T_s$ in FeSe, as evident from the band splitting around the M point. The present result shows that the long-range magnetic order is not a prerequisite to induce the sizable band splitting. We also found that this band splitting leads to the appearance of a van Hove singularity near $E_{\rm F}$. Our result suggests that such unconventional electronic states are responsible for the anisotropic superconducting states in FeSe which exhibit possible gap nodes.

\begin{acknowledgements}
We thank H. Kumigashira and K. Ono for their assistance in ARPES measurements. We also thank J. Kleeman for his critical reading of our manuscript. This work was supported by grants from the Japan Society for the Promotion of Science (JSPS), the Ministry of Education, Culture, Sports, Science and Technology (MEXT) of Japan, and High Energy Accelerator Research Organization, Photon Factory (KEK-PF) (Proposal No. 2012S2-001).
\end{acknowledgements}

\bibliographystyle{prsty}

\begin{thebibliography}{30}
\bibitem{note} Since the nematicity and the orthorhombicity commonly have the C$_2$ symmetry and therefore easily couple with each other, the development of nematicity would induce the structural instability \cite{FernandesReview}.
\end{thebibliography}

\end{document}